# Transaction Confirmation Time Prediction in Ethereum Blockchain Using Machine Learning

Harsh Jot Singh, Abdelhakim Senhaji Hafid, Department of Computer Science and Operational Research, University of Montreal, Montreal, QC H3T 1J4

*Abstract-* **Blockchain offers a decentralized, immutable, transparent system of records. It offers a peer-to-peer network of nodes with no centralised governing entity making it 'unhackable' and therefore, more secure than the traditional paper-based or centralised system of records like banks etc. While there are certain advantages to the paper-based recording approach, it does not work well with digital relationships where the data is in constant flux. Unlike traditional channels, governed by centralized entities, blockchain offers its users a certain level of anonymity by providing capabilities to interact without disclosing their personal identities and allows them to build trust without a third-party governing entity. Due to the aforementioned characteristics of blockchain, more and more users around the globe are inclined towards making a digital transaction via blockchain than via rudimentary channels. Therefore, there is a dire need for us to gain insight on how these transactions are processed by the blockchain and how much time it may take for a peer to confirm a transaction and add it to the blockchain network. This paper presents a novel approach that would allow one to estimate the time, in block time or otherwise, it would take for a mining node to accept and confirm a transaction to a block using machine learning. The paper also aims to compare the predictive accuracy of two machine learning regression models- Random Forest Regressor and Multilayer Perceptron against previously proposed statistical regression model under a set evaluation criterion. The objective is to determine whether machine learning offers a more accurate predictive model than conventional statistical models. The proposed model results in improved accuracy in prediction.**

**Index terms-- Blockchain, Confirmation Time, Ethereum, Machine Learning, Regression, Transaction, Multilayer Perceptron (MLP), Random Forest.**

## I. INTRODUCTION

Blockchain Technology is a distributed database shared between nodes in a peer-to-peer network (e.g., more than 10000 nodes in Ethereum). Basically, each network node can receive and broadcast transactions. Blockchain, as the name suggests, records transactions into linked blocks [1]. When a user wants to interact with the blockchain (e.g., to transfer cryptocurrency or store a testament), they create and sign, using their private key, a transaction; note that blockchain, in itself, uses public key encryption. Then, it sends the transaction to the blockchain network; a node that receives the transaction, validates the transactions (e.g., verifies the user's signature) and, if valid, stores the transaction in its pending list of transactions and transmits it to its neighbouring nodes. Periodically, a node is selected to create a block; the selection is based on the consensus protocol in use. In the case of proof-of-work (PoW) consensus protocol [2], the node that first solves a mathematical puzzle, is the one that creates the new block. It is important to emphasize that there should be no shortcuts to solve the puzzle in order to guarantee that nodes are selected randomly. PoW consists of determining a string (called nonce) such that when combined with the block header and hashed results in hash that includes a given number of leading 0 bits (this number represents the difficulty in solving the puzzle). Nodes are incentivised to create new blocks because they are rewarded by newly minted coins (e.g., in the bitcoin blockchain, the reward is 12.5 bitcoins as of 2019) and transactions fees. The time it takes to generate a block, called block time, is specific to the blockchain in use; for example, the block time for bitcoin is 10 minutes whereas it is 15 seconds for Ethereum.

Blockchain comes in many different types. More specifically, there are three types of blockchains: permissionless blockchain also known as public blockchain (e.g., Bitcoin and Ethereum), permissioned blockchain also known as consortium blockchain (e.g., Hyperledger fabric), and private blockchain. In public blockchains, any participant/user can write data to the blockchain and can read data recorded in the blockchain; anybody can be a full node, a miner or a light node. Thus, there is little to no privacy for recorded data and there are no regulations or rules for participants to join the network. Generally, pubic blockchains are considered pseudo-anonymous (e.g., bitcoin and Ethereum); a participant does not have to divulge their identity (e.g., name) instead the user is linked to an address (i.e., hash of public key). Providing anonymity is difficult but it is feasible (e.g., Zcash [3]). The success of this type of blockchains depends on the number of participants; it uses incentives to encourage more participation. Consortium blockchains put restrictions on who can participate. In particular, the creation and validation of blocks are controlled by a set of pre-authorized nodes; for example, we have a consortium of 10 banks where each bank operates one node. The right to read data recorded in the blockchain can be public or restricted to the participants. Even participants may be restricted on what they can do in the blockchain; for example, transactions between 2 participants may be hidden from the rest of participants. In a private blockchain, write permissions are centralized and restricted to one entity; read permissions may be public or restricted.

Blockchain offers a way for users to exchange value with all the capabilities offered by state-backed currencies but is more secure and does not require central governing entity. The biggest application of blockchain is cryptocurrency. Cryptocurrency is a digital or virtual type of currency that uses encryption techniques to convert plain text into unintelligible text and vice-versa. It is designed to work as a medium of exchange as well as to control the creation of additional units and validate transactions and allow their transfer through the blockchain network. Bitcoin, Namecoin, Ethereum etc. are a few examples of most commonly used cryptocurrencies.

The market capitalization of publicly traded cryptocurrencies as of October 30, 2019, is about 241.9 billion out of which Bitcoin makes up to 157.4 billion and Ethereum makes up to 26.6 billion dollars [4]. On average, about 12 billion USD is transferred via at least a million transactions each day. The amount of capital involved alone makes it necessary for one to be able to perform analysis on historical and real-time data and allow one to infer and/or predict a certain level of details about future trends in the market. These trends can range from the analysis of address space [5], price prediction [6], transaction confirmation time prediction [7][8], etc.

Bitcoin, when it first came to be in 2009, was not simply meant to be a tool to spend money digitally. It was meant to be a convergence of networking, cryptography and open source software technologies with the aim to completely eradicate the need of state-backed currencies by crossing international boundaries and nullifying the usage of banks as a mean to store money [9]. Since then, there has been a gradual increase in the awareness and excitement in regard to cryptocurrencies and the rudimentary distributed ledger (or Blockchain) technology. On the grassroot level, these blockchain based cryptocurrencies are meant to provide complete anonymity (or rather pseudo-anonymity) to its users by providing them capabilities to operate via a set of addresses without having to disclose any of their personal details [5] all the while providing high level of transparency on past transactions.

Ethereum, the second most commonly used cryptocurrency was launched in 2015, is the most well-established, largest open-ended, public and blockchain-based software platform. Unlike Bitcoin, which only allows for value exchange, Ethereum permits the utilization of Smart Contracts. These are snippets of code or protocols that digitally facilitate, verify and ensure performance of a contract [10]. It also offers a set of programming languages allowing users to publish Distributed Applications without external interference, fraud or downtime using its own decentralized public blockchain technology [11] [12].

Ethereum is a decentralized technology. It runs on a network of machines (or nodes) that are distributed globally. Since there is no central point of failure, Ethereum is also immune to hacking or any attack preventing its operation. These characteristics of Ethereum allow it to be more formidable in comparison to its counterparts and hence, make it possible for Ethereum data to be used for knowledge inference and/or analysis. As such, in this paper, we particularly focus on Ethereum and the time it takes for a mining node to confirm a transaction on the said platform. A transaction is how the external world interacts with the Ethereum network. Each time there is a need to modify or update the state of the network, a transaction has to me made. These transactions can be of three types: (1) Fund transfer between two accounts; (2) Deployment of a contract on the Ethereum network; and (3) Execution of a function on a deployed contract.

Any change in the state of the network is considered to be a transaction. The transaction carries information of the user, via the user interface (e.g., browser), to the network, to another endpoint on the network or back to the user's station. It could carry large amount of capital in form of ether or data through contracts that a transaction can call for execution.

Due to the importance of these transactions and the amount of capital these might carry within them, it is very important for a user to gain some insight on how much time it might take for the transaction to be processed based on the network traffic. By gaining these insights ahead of time, a user can infer whether right now would be the right time for them to send this transaction to the network. Not only this, by understanding how much time the miners will take to process a transaction, the user can gain insight on miner policies, i.e., what factors are taken into consideration by mining nodes while choosing one transaction over the others. In general, a miner would choose a transaction where they would get the most incentive. But, since there is no set policy for the Ethereum blockchain, these policies can change at any time and by analysing the most recent network trend, the user can keep up to date with these policies.

Each time a user makes a transaction, they have to pay a fee. Again, while miners will process transactions with higher fees first, it is not efficient for a user to send a transaction with a value so low that the transaction would never be picked up and they would have to resend the transaction at a higher value. Similarly, it would not be helpful (or rather it would be wasteful) for the user to make a transaction with fees higher than what miners are accepting to prioritize transactions. Gaining insight on the current network state will assist the user to determine "optimal" fees they need to pay for her transaction. The paper employs two machine learning algorithms, i.e., Multi-Layer Perceptron and Random Forest to make these predictions. The paper also aims to compare the performance of the two models as well as the performance of previously employed statistical predictive models [8] to determine which model presents the best predictive accuracy when it comes to prediction of Ethereum blockchain confirmation time.

## II. ETHEREUM

Ethereum was developed to facilitate transactions among individuals without requiring them to disclose their personal details, i.e., by developing trust among two anonymous entities. As a whole, it is a transaction-based state machine beginning at the genesis state [13] and moves on to the final state via incremental execution of transactions [14]. In between the two states, there can be valid or invalid changes. The invalid state changes may or may not imply an invalid account balance modification in either the sender's or the receiver's account. Formally [2],

$$\sigma_{t+1} = Y(\sigma_t, T) \quad (1)$$

where, $Y$ is Ethereum state transition function allowing components to carry out computations and $\sigma$ allows them to store arbitrary state between transactions.

These transactions are accumulated into blocks by utilizing Merkle trees [15]. These blocks work as journals recording the transactions and are connected to one another using a cryptographic hash. Blocks, instead of storing the final state, only store an identifier to it. This is because storing the final state would require far more storage than what is available within a block. An Ethereum block is a collection of the block header ($B_H$), all the data relevant to comprised transactions ($B_T$) and a set of other block headers ($B_U$) that are known to have a parent block equal to the current block's parents' parents, known as ommers. Formally,

$$B \equiv (B_H, B_T, B_U) \quad (2)$$

Blocks also accentuate transactions with incentives for the mining nodes. This incentivization is a state-based function that adds value to a nominated account [2].

Mining is the process of emphasizing on one set of transaction series or block over others. It is achieved via a cryptographically secure process called 'proof-of-work'. Formally [2],

$$\sigma_{t+1} \equiv \Pi(\sigma_t, B) \quad (3)$$

$$B \equiv (\ldots, (T_0, T_1, \ldots)) \quad (4)$$

$$\Pi(\sigma, B) \equiv \Omega(B, Y(Y(\sigma, T_0), T_1)\ldots) \quad (5)$$

where Ω is the block-finalisation state transition function used to reward the nominated account; B is the block in question and Π is the block-level state function.

This represents the basic blockchain paradigm on which Ethereum and all blockchain based technologies operate on.

*A. Accounts*

The basic unit of Ethereum is account. Each account has a 20-byte address. For an individual to make a transaction they need to have an account. These Ethereum accounts are of two types: Externally Owned Account (EOA) or Contract Accounts.

EOA are controlled by private keys and have ether balance. They are capable of sending transactions to the blockchain. These transactions can be to transfer Ether (i.e., Ethereum cryptocurrency), trigger contract accounts or update the state of the machine. There is no code associated with externally owned accounts whereas contract accounts always have some code associated with them. The code execution is triggered by transactions or message calls received from other contracts or EOA. Whenever a contract account is executed, it performs operations of arbitrary complexities, manipulates its own persistent storage or its own permanent state and/or call other contracts.

*B. Transactions*

A transaction is how the external world interacts with the Ethereum platform. It is a cryptographically signed instruction sent by an account holder to change the state of Ethereum at any given time. There are two types of transactions in Ethereum: (1) Transactions made to generate message calls or (2) To create new accounts with code, i.e., contract accounts. These transactions have some components associated with them [2]. These components are discussed in Table I.

TABLE I: COMPONENTS OF A TRANSACTION IN ETHEREUM

| Components | |
|---|---|
| nonce | Represents the number of transactions sent by the sender ($T_n$). |
| gasPrice | Amount of Wei (see Table II) to be paid per unit of gas for the computation cost incurred due to the execution of a transaction ($T_p$). |
| gasLimit | Represents the maximum amount of gas that can be used to execute a transaction. This is paid upfront and cannot be modified later ($T_g$). |
| to | 160-bit receiver's address or that of a contract creation transaction ($T_t$). |
| value | Amount of Wei to be transferred ($T_v$). |
| v, r, s | Transaction signature; it is also used to determine the transaction source or sender ($T_v, T_r$ and $T_s$). |
| init | Associated with contract creation transactions; it is an unlimited size byte array that specifies the EVM-code for account initialization procedure ($T_i$). It is an EVM code fragment that returns the *body*, the second code fragment to be executed each time the account receives a message call while *init* is executed only once. |
| data | Associated with message call transactions; it is an unlimited sized byte array containing the input data of the message call ($T_d$). |

TABLE II: DENOMINATION OF ETHER

| Multiplier | Name |
|---|---|
| $10^0$ | Wei |
| $10^{12}$ | Szabo |
| $10^{15}$ | Finney |
| $10^{18}$ | Ether |

## C. Gas and Payment

In order to prevent network abuse, all transactions on Ethereum are subject to a fee [2]. This fee is paid in 'gas'. Gas is the fundamental unit used to measure the cost of computation and execution of a transaction on Ethereum.

Every transaction has a specified amount of gas associated with it known as gasLimit (see Table I). This is the prepaid amount that is charged to the sender account before the transactions is processed. The amount is charged based on another entity included in a transaction called gasPrice. For instance, if a user sets the gas limit to 40,000 and a gas price of 25 gwei. This would imply that the user is willing to spend at most 40,000 x 25 gwei or $10^{-3}$ Ether in order to execute the transaction. If an account cannot pay for a transaction it wants to process, the transaction would be considered failed or invalid. On the other hand, if a transaction requires less amount of gas than what is paid by the account then the remaining gas is refunded back to the diner's account.

## D. Transaction Execution

When a user wishes to make an exchange or execute a contract, initially, the data would have to be converted into a raw transaction data. This raw transaction contains the gasPrice, the gasLimit set by the user, the destination addresses as well as value; which is the total amount of Ether they wish to send. It would then include the data into it which would be the hash of the function in the contract that the account holder wishes to execute.

In order to ensure that the transaction is made by the account holder and not by an unauthorized entity, the transaction would then have to be signed with the user's private key. This is also used to ensure accountability of the sender.

The transaction is then broadcasted to the Ethereum network and the local geth (or parity) node will generate a transaction id that the user will be able to use to track its status. These geth (GO-Ethereum) or parity (Parity-Ethereum) nodes are some Ethereum protocols defining how a user would operate, how the network works and rules each user must follow to be a valid part of the Ethereum network [16]. Figure 1 illustrates the broadcasting process.

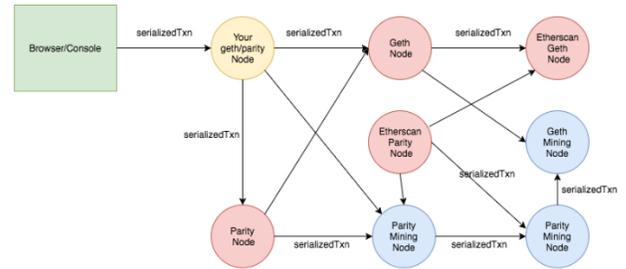

Figure 1: Signed Transaction propagating through the network. The transaction is broadcasted by sender's local geth (or parity) node to its peers who then broadcast it to their peers and so on until the whole network has a signed copy of the transaction.

In the Ethereum network, some nodes work as full nodes or mining nodes. These nodes pick up a transaction and put in the effort to include the said transaction into a block. Mining nodes have a transaction pool where each transaction exist as pending before it is picked up for evaluation. These transactions are stored in the pool as per the gas price associated with each one of them. Higher the price for a transaction, more likely is the node to evaluate it first. A sample pending pool is illustrated in Figure 2.

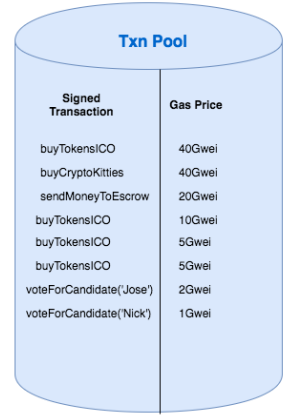

Figure 2: Mining node's pending transaction pool.

It should be noted that the pending transaction pool can only hold a limited number of transactions. If account holders keep on generating transactions with higher gas price, as is the trend, the transaction with the minimum gas would be discarded and the transaction would have to be re-broadcasted to the network.

Next, once the miner has picked up the transaction to include in the block, the transaction would be validated, included into the pending block and the proof of work begins. Then, once the block is added to the blockchain, the valid block is then broadcasted to the entire network by the mining node.

Finally, the local node will receive the new block and will then sync its local copy of the blockchain. It would then execute all the validated transactions in the block.

### III. RELATED WORK

Due to the volatile nature of the Ethereum transaction dataset, its predictability has been sparsely covered in published literature. However, this section briefly discusses some of the work involving Ethereum transaction confirmation time prediction and their limitations.

Singh and Hafid [7] proposed a prediction model by considering it as a classification problem. The proposed model splits transaction confirmation time into eight predefined classes. The model then classifies new transactions based on what is learnt from the historical as well as the real time data. The paper employs weighted Naïve Bayes, weighted Random Forest and MLP models. The three models work with an average accuracy of about 83.5%. Their model splits the confirmation time of transaction into eight classes: within 15 seconds, within 30 seconds, within 1 minute, within 2 minutes, within 5 minutes, within 10 minutes, within 15 minutes and within 30 minutes or longer. We know that on average, a transaction has to wait for two block confirmations (~30 seconds) before it is confirmed. However, in cases where the model would predict that the transaction belongs to 'within 5 minutes' class, there is no way for the user to know if it would take 3 minutes, 4 minutes or more. Hence, while the paper presents models with good prediction accuracy, it considers confirmation time prediction as a simple classification problem. It can only provide a user with an approximation of time it would take for their transaction to be confirmed which may or may not always be ideal.

Unlike Singh and Hafid's classification model, Eth Gas Station [8] proposed a Poisson regression model to estimate the expected number of blocks it would take for a transaction to be confirmed based on the amount of gas used by the transaction and the gas price. The model outputs real values (in seconds and in blocks) instead of generalizing to a category like in Singh's model. The model makes its estimations based on the data from the last 10,000 blocks at any given time. The model uses statistical analysis and outputs the confirmation time prediction based on the percent of blocks that had a similar transaction confirmed within them in the past. The Poisson regression model proposed by them is periodically retrained to keep up with the most recent state of the Ethereum network. The model does not take into consideration the transactions in the past that are still pending and how the data from those transactions could relate to current transactions. Their proposed statistical model is used as the basis of this paper. It is used to analyse which of the two, machine learning or statistical modelling, would be more efficient for the prediction problem.

### IV. PROPOSED MODEL

In this section, we present the proposed method for the prediction problem. We first discuss the dataset and the associated features as well as how it was utilized to solve the problem. We then briefly discuss the machine learning models used in order to predict the confirmation time for a transaction.

*A. Dataset and Features*

The initial dataset consists of about one million transactions that were made on Ethereum with the last transaction in the set dating on November 18, 2018. The data was extracted from ethereum.io API [17] and included the gas price, gas limit set by the user, gas used, the timestamp for when the transaction was confirmed and when the transaction was made as well as the number of transactions made by the sender. While there is no given field in the API that would suggest when the transaction was made, a python script to extract the timestamp from the pending transaction pool [18] was used. These transactions were used as the historical data to train the model before the actual time estimation was made. While about 70 % of these transactions were confirmed within one or two block confirmations, about 1% were also reverted or failed (see Table III). We wrote a python script with the purpose of monitoring a local Geth node. This was done to extract data about mined as well as pending transactions. This allowed the model to access real-time data and make predictions based on current market trends rather than simply relying on historical data.

TABLE III: STATISTICS OF THE DATASET

| Confirmation Time (15 sec ≡ 1 block) | Number of Transactions |
|---|---|
| 1. one block | 494,071 |
| 2. two blocks | 248,609 |
| 3. four blocks | 173,047 |
| 4. six blocks or longer | 113,673 |

In order to predict discrete valued output defining the time taken by a mining node to confirm a transaction for our regression algorithms, the independent are simply passed through a regressor to determine the dependent variable, conf_time. This dependent is the difference between the two timestamps: (1) transaction's introduction to the network and (2) the timestamp for its confirmation. Table IV lists the variables in the input and output vector for our machine learning model.

It should be noted that while on average, it takes up to 2 blocks for a transaction to be confirmed, it varies depending on network traffic as well as the number of transactions that a mining node is assigning to a block. Another parameter that affects the confirmation time for a transaction is the gas limit the user has imposed on the transaction in question. A mining node gets an incentive for working on each transaction (see Section I.A.). Higher the limit imposed by the user, the higher the incentive the miner would get for the effort put in to perform computations on the transaction. Therefore, the mining node would be inclined to confirming such transactions over others leading to lower confirmation time. The variation in the number of pending transactions per minute can be observed in Figure 3 [19] and the block confirmation time variation can be observed in Figure 4 [20].

TABLE IV: DEPENDENT AND INDEPENDENT VARIABLES OF THE DATASET

| Variable | Type | Description |
|---|---|---|
| gasPrice | Independent | Gas price provided by the sender in Wei. |
| gasLimit | Independent | Represents the maximum amount of gas that should be used to execute a transaction. |
| value | Independent | Value transferred in Wei |
| receipt_status | Independent | Either 1 (success) or 0 (failure). This reports whether the transaction was successfully confirmed or not. |
| timestamp_0 | | Timestamp for when the transaction was added to the pending pool. |
| timestamp_1 | | Timestamp for when the transaction was confirmed by the mining node. |
| conf_time | Dependent | Difference between timestamp_0 and timestamp_1. This is used as the dependent variable in case of regression. |

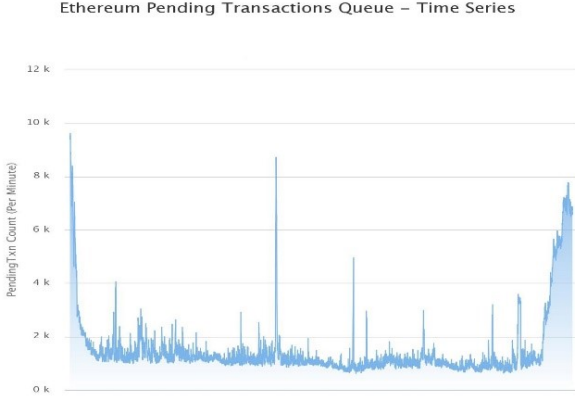

Figure 3: Ethereum pending transaction queue (per min.)

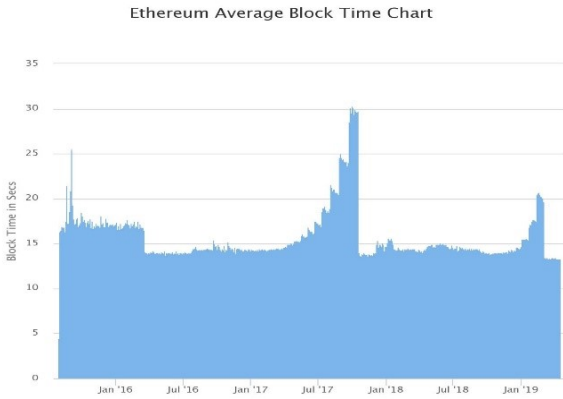

Figure 4: Ethereum Average Block time chart. (in seconds)

## B. Machine Learning Models

One of the most important steps in data analysis using machine learning is choosing the most appropriate algorithm for the dataset. While there are many regression algorithms available, the performance of these different techniques largely depends on the structure and the size of the dataset. In this paper, we employed Random Forest and MLP regressors to solve the prediction problem. The models and the rationale behind choosing them is discussed as follows.

### 1) Random Forest

A random forest is an ensemble machine learning technique that can perform both regression and classification tasks by utilizing multiple decision and Bootstrap Bagging [21]. The fundamental idea behind bagging is to create random noise in the dataset. This is done by making multiple copies of the data each with a slight variation from the original. It should be noted that while there is a certain amount of noise involved in bagging, the model is considered to be unbiased with an aim to reduce variance by averaging the noise in each of the varied instances. Random forest uses multiple decision trees and then outputs the mean prediction of each tree, in case of regression, and the class that is the mode of the classes in case of classification. Random forest approximates the expectation as follows:

$$\hat{f}_{rf} = E_\theta T(x; \varphi) = \lim_{Y \to \infty} \hat{f}(x)_{rf}^Y \qquad (6)$$

with an average over $Y$ realizations of $\varphi$. Figure 5 shows an instance of a simple random forest.

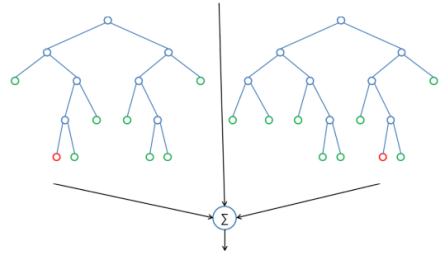

Figure 5 A simple random forest with three trees

Random forest model, in general, performs well at learning complex, highly non-linear relationships; like between time and both the gas price and the gas used in Ethereum blockchain dataset. The model is known to outperform fundamental classification and regression models like naïve Bayes, polynomial and linear regressors [22]. The model proposed in the paper employs random forest regressor to make confirmation time predictions.

### 2) Multi-Layer Perceptron (MLP)

MLP is an adaptive neural structure made up of at least three layers constituting an input layer; which is made up of a number of perceptions equal to the number of attributes of the dataset. The network must have at least one hidden layer and an output layer made up of one perceptron in case of regression and a number of perceptrons equal to the number of classes in case of a classification problem. These layers communicate among one another via synaptic connections represented by weights [23]. Each layer except the input layer has a non-linear activation function.

MLP is a supervised learning technique which learns a function $f(\cdot): R^n \to R^o$ by learning a dataset with $n$ input dimensions and $o$ output dimensions. Figure 6 shows a simple MLP regressor.

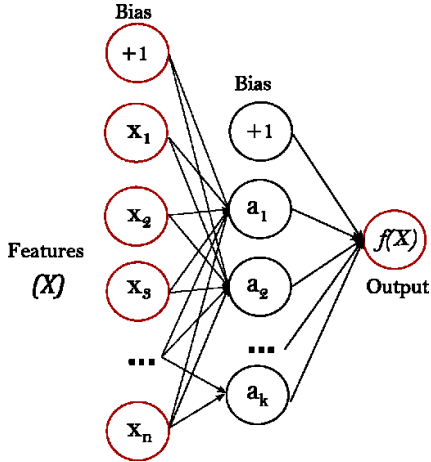

Figure 6 An instance of MLP regressor with one hidden layer.

An MLP can work with as many hidden layers and parameters as one might require, all with non-linearities between them. This allows the model to be able to learn any complex non-linear relationship in a dataset, as observed with random forest. MLP is known to be flexible and adaptive to any feature-variable relationships and hence, does not require one to stress over the structure of the network [23].

The proposed model employs an MLP regressor that trains using backpropagation and no activation function (or identity function) at the output layer. It employs square error as the loss function; it is expressed as follows:

$$Loss(\hat{y}, y, W) = \frac{1}{2}\|\hat{y} - y\|_2^2 + \frac{\alpha}{2}\|W\|_2^2 \qquad (7)$$

where $y$ represents the predicted output of the model, $\hat{y}$ represents the actual output, $\alpha\|W\|_2^2$ is the L2-regularization that penalizes models and $\alpha > 0$ is a non-negative hyperparameter controlling the magnitude of penalty. MLP regressor, starting from random weights $W$, works on minimizing the aforementioned loss function by running multiple iterations and updating the weights at every step as it backpropagates through the output layer. The output is a set of continuous values, i.e., the confirmation time predictions for the test data.

## V. EVALUATION CRITERION AND SETUP

Before we access the performance of any given model; especially in case of regression where the output is a set of continuous values and not a set of discrete values, it is necessary to have a set criterion for evaluation. In the following section, we discuss the complexity of the two models and compare it to the complexity of the statistical model used by Eth Gas Station [8]. The section also discusses the evaluation criteria employed to measure the performance of the two proposed models against one another and the statistical model [8]; then, it presents the experimental setup used for the performance measurement.

### A. Computation Complexities

The computational complexity of an algorithm can be defined as the measure of resources required for running it. For a given instance or input vector [24], it is the measure of the number of steps required to compute the output, assuming this to be the worst-case scenario. In general, the complexity of machine learning algorithms is highly dependent on their implementations, dataset properties, etc. [25]. While discussing the complexity of the two machine learning algorithms employed in this paper, we consider upper bounds in case of dense data as is the case with our dataset.

Let $n$ denote the number of instances or transactions in our training subset, $p$ denote the number of parameters passed as input vector (seven in our case), $h_{li}$ denote the number of nodes in each layer $i$ of MLP, $e$ denote the number of epochs and $t$ denote the number of trees in the random forest method. Table V shows the complexities of the three approaches.

TABLE V: COMPLEXITY OF THE THREE ALGORITHMS

| Model | Training | Prediction |
|---|---|---|
| MLP | $O(e*n*(ph_{l1} + h_{l1}h_{l2} + ...))$ | $O(ph_{l1} + h_{l1}h_{l2} + ...)$ |
| Random Forest | $O(n^2 pt)$ | $O(pt)$ |
| Eth Gas Station | $O(p^2 n + n^3)$ | $O(p)$ |

We observe that random forest, which is an ensemble method, simply multiplies the prediction accuracy, of a simple regressor used by Eth Gas Station, by the number of trees or 'voters' employed. While one interpretation of this could be increased complexity, the presence of multiple voters also ensures a stronger model with a chance at better predictive accuracy.

Similar inference can be made in case of MLP. The complexity of an MLP depends on the number of layers, the number of neurons in each layer as well as the number of training epochs. While the network structure may increase, it also leads to a stronger more accurate predictive model which is not the case when we consider a liner regressor like the one used by Eth Gas Station.

### B. Evaluation Criterion

The evaluation consists of comparing the time predicted by the proposed model against the time it actually took for a transaction to get confirmed by a mining node as per the data collected by monitoring the local Geth node. The models were evaluated on the following criteria:

#### 1) Mean Absolute Error

Mean Absolute Error is defined as the mean of absolute error or difference between predicted and actual values, that is, predicted confirmation time and the actual confirmation time. Mathematically, this can be defined as follows:

$$MAE = \frac{1}{n}\sum_{i=1}^{n}|act_i - est_i| \quad (8)$$

#### 2) Root Mean Square Error

RMSE represents the standard deviation of the magnitude difference between predicted and actual values. It measures the square root of the average of the squared difference between the predicted value and the true values. Formally,

$$RMSE = \sqrt{\frac{1}{n}\sum_{i=1}^{n}(act_i - est_i)^2} \quad (9)$$

#### 3) Mean Absolute Percentage Error

MAPE is percentage equivalent of MAE. It is most commonly used as a loss function in regression problems and model evaluation due to its intuitive interpretation of relative error. Formally,

$$MAPE = \frac{1}{n}\sum_{i=1}^{n}\left|\frac{act_i - est_i}{act_i}\right| \times 100\% \quad (10)$$

#### 4) Prediction Value (PRED)

PRED(n) represents the percentage of absolute percentage error that is less than or equal to the value n among N transactions.

$$PRED(N) = \frac{1}{n}\sum_{1}^{n}\begin{cases}1, \text{if MAE} \leq N \\ 0, \text{otherwise}\end{cases} \quad (11)$$

### C. Experimental Setup

In our experiment, we trained multiple variations of MLP with one hidden layer, two hidden layers as well as with three hidden layers to compare the effect of the added layer on training complexity and the prediction accuracy of the model. Similarly, random forest model with 250 and 500 trees and at least 5 samples at leaf nodes, with 1000 trees and at least 10 samples, 1500 trees and at least 15 samples as well as one with 2000 trees and at least 20 samples at leaf nodes was implemented.

For the dataset, we initially split the historical data into two parts where 80% of the transactions were fed to both of our proposed models, i.e., MLP and random forest, and the remaining 20% was used to validate the models after the training phase. This was done to make sure that the model was able to properly predict the block confirmation time for the transaction before real-time data was involved.

The computational complexity of MLP is dependent on the number of hidden layers used to design the network and that of the random forest depends mostly on the number of trees used (see Section IV). The same was observed when we executed the aforementioned variations of the two machine learning models the results of which can be observed in Table VI, VII and Figures 7 and 8.

TABLE VI: TRAINING TIME FOR THE THREE VARIANTS OF MLP

| Model | Time to Train (in minutes) |
|---|---|
| MLP with one hidden layer | 130 |
| MLP with two hidden layers | 210 |
| MLP with three hidden layers | 350 |

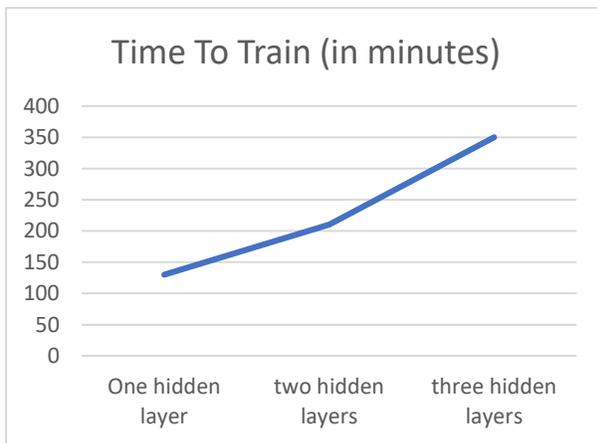

Figure 7: Training time for different variants of MLP (in minutes).

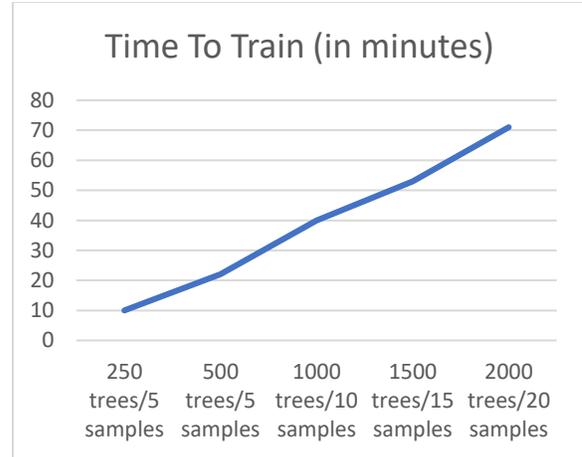

Figure 8: Training time for different variants of Random Forest (in minutes).

TABLE VII: TRAINING TIME FOR THE FIVE VARIANTS OF RANDOM FOREST

| Model | Time to Train (in minutes) |
|---|---|
| 250 trees/5 samples | 10 |
| 500 trees/5 samples | 22 |
| 1000 trees/10 samples | 40 |
| 1500 trees/15 samples | 53 |
| 2000 trees/20 samples | 71 |

We also observed that there was a gradual increase in prediction accuracy of the model as we added more layers, in case of MLP and more trees, in case of random forest. However, the increase was so slight that when considering the time taken to train these variations, the increased model accuracy turns out to be a bad trade-off. The performance of the two models and the variants can be observed in Table VIII, IX and Figures 9 and 10.

TABLE VIII: RESULT ANALYSIS OF STATIC DATASET FOR THE THREE VARIANTS OF MLP

| Model | MAE | RMSE | MAPE | Pred (0.20) |
|---|---|---|---|---|
| One hidden layer | 0.167 | 0.408 | 7.40% | 87.51% |
| Two hidden layers | 0.153 | 0.391 | 6.89% | 88.02% |
| Three hidden layers | 0.1493 | 0.386 | 6.57% | 88.44% |

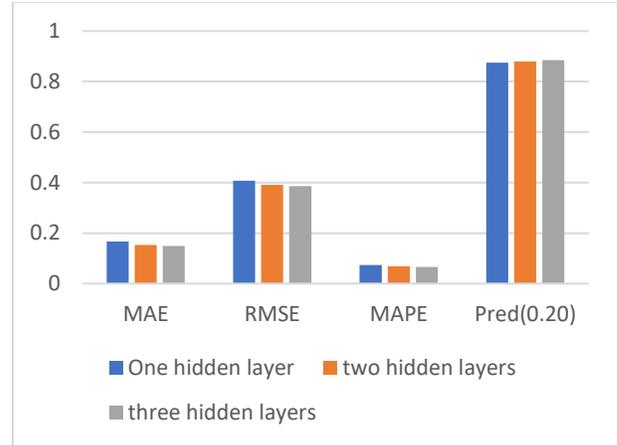

Figure 9: Performance comparison between variations of MLP, with static data and under the set evaluation criteria.

TABLE IX: RESULT ANALYSIS OF STATIC DATASET FOR THE FIVE VARIANTS OF RANDOM FOREST

| Model | MAE | RMSE | MAPE | Pred (0.20) |
|---|---|---|---|---|
| 250 trees/5 samples | 0.118 | 0.343 | 5.07% | 89.54% |
| 500 trees/5 samples | 0.112 | 0.334 | 4.83% | 93.79% |
| 1000 trees/10 samples | 0.109 | 0.33 | 4.81% | 93.91% |
| 1500 trees/15 samples | 0.1087 | 0.329 | 4.809% | 93.92% |
| 2000 trees/20 samples | 0.1087 | 0.329 | 4.809% | 93.92% |

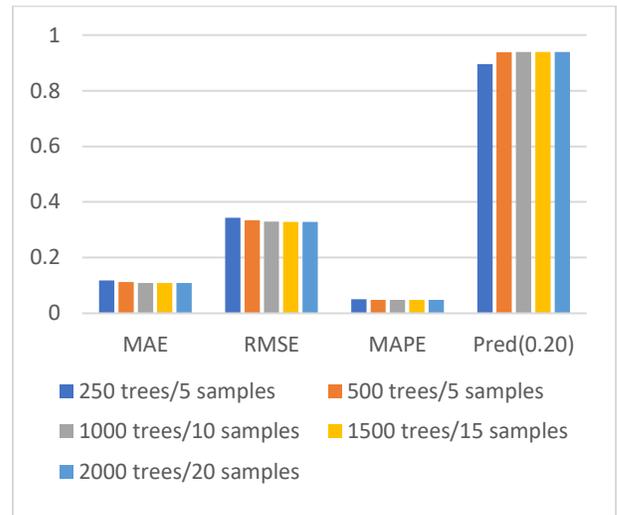

Figure 10: Performance comparison between variations of, Random Forest with static data and under the set evaluation criteria.

For real-time data, the local geth node that was being monitored (see Section III) has been recording details of the transactions as they are made. These details include the timestamp for when the transaction was made, how many blocks were created before it was confirmed, the gas price as well as the gas limit set by the user. Whenever there is a set of new transactions for which the model is required to make confirmation time predictions, it would do so based on the data collected by geth node for the most recent confirmed transactions (from the past 100 blocks). These are then fed to the model with the main strategy of predicting the confirmation time for a transaction given a certain gas price between 0-100 gwei at the current state of the transaction pool.

## VI. RESULT ANALYSIS AND DISCUSSION

The two machine learning regression techniques proposed were compared with Poisson Regression Statistical model used by Eth Gas Station [8]. Note that there is no literature available confirming the accuracy of this statistical model. To measure the prediction accuracy of Eth Gas Station, a randomly generated subset of the data gathered from the local geth node was passed through a regression model similar to the one used by Eth Gas Station as available on their GitHub page [26]. Eth Gas Station performs their predictions in terms of number of blocks a transaction had to wait before it was confirmed by a mining node instead of number of seconds as done in this work [27]. It should be noted that, on average, in Ethereum, a new block is created every 15 seconds. This implies that if a transaction had to wait 29 seconds before it was confirmed by a mining node then it waited for two block confirmations. This interchangeability between the two allowed us to compare the performance of our machine learning algorithms against Eth Gas Stations Poisson regression model.

According to the data collected from BitInfoCharts [28], on average, each block has 138 transactions in it. Considering that our models consider data from 100 most recent blocks to predict the confirmation time for the transaction(s) in question, the models will have to learn about 13.8 thousand transaction. Table X illustrates the time taken by each model to learn these transactions.

TABLE X: TRAINING TIME FOR THE THREE MODEL TO LEARN 100 BLOCKS

| Model | Time to Train (in minutes) |
| --- | --- |
| Eth Gas Station | 2.91 |
| MLP | 7.31 |
| Random Forest | 2.78 |

Taking into consideration the complexity of each variant as observed in the previous section, comparison of statistical approach taken by Eth Gas Station was done against MLP with one hidden layer and random forest with 500 trees and at least 5 samples at leaf nodes. It should be noted that for Eth Gas Station, we implemented a Poisson regression model to calculate the time taken by the model to learn 13.8k transactions.

Considering that a user might want to make predictions at any given time and the time taken by each model to learn most recent data, models need to be updated at all times. We, therefore, trained the random forest model at an interval of 3 minutes whereas the MLP was trained at an interval of 8 minutes. Table XI and figure 11 shows the results for the two models when compared with Eth Gas Station.

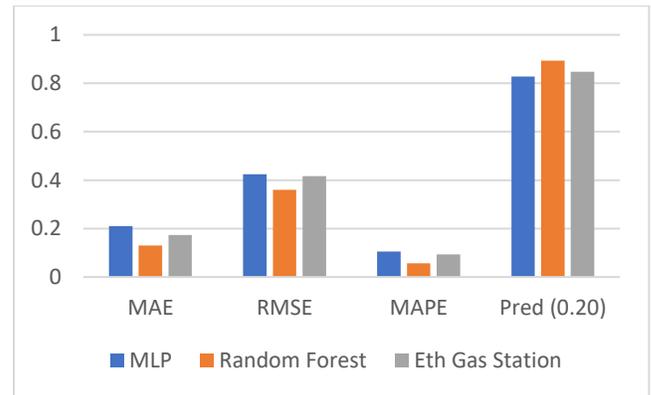

Figure 11: Performance comparison between MLP, Random Forest and Eth Gas Station under the set evaluation criteria.

TABLE XI: RESULT ANALYSIS FOR THREE MODELS

| Model | MAE | RMSE | MAPE | Pred (0.20) |
| --- | --- | --- | --- | --- |
| MLP | 0.21 | 0.424 | 10.54% | 82.74% |
| Random Forest | 0.13 | 0.36 | 5.70% | 89.36% |
| Eth Gas Station | 0.174 | 0.417 | 9.31% | 84.79% |

The Multi-Layer Perceptron produces a mean absolute error of about 0.21 with the root mean square error of 0.424 and pred(n) of 82.74% at n being 0.20. While analyzing hyperparameters, it was observed that RMSE can be decreased at the cost of prediction accuracy, resulting in a weak or inaccurate model. The mean absolute percentage error observed with MLP was about 10.54%. Similarly, with random forest, MAE was observed to be good at 0.13 with RMSE at 0.36, MAPE at 5.7% and pred (0.20) of a solid 89.36%. When evaluating the statistical Eth Gas Station, the mean absolute error of 0.174 was observed with the mean absolute percentage error at 0.31%; RMSE at 0.417 and the prediction accuracy of 84.79%. We conclude, from the above discussion, that the random forest model performs better than the other two models in all the evaluation criterions employed.

## VII. CONCLUSION

The paper systematically compares the performance of two machine learning regression models and the more classical, statistical model on the task of predicting the confirmation time for a transaction in Ethereum Blockchain. Due to the volatility of the Ethereum Blockchain, as seen in the changes to the network due to congestion, variation in gas prices; the prediction of confirmation time for a transaction demands that prediction models keep up with the most recent dataset instead of relying on historical data. We discussed multi-layer perceptron and random forest regressors; the two machine learning techniques employed in this paper and their variants. We observed that while there is an increase in the prediction accuracy of the two algorithms under the more complex variants, the trade-off between the better accuracy and the training complexity of the said variants is unnecessary considering how slight the difference from a simpler variant. The results discussed in section V suggest that machine learning, which prior to this work was never employed for such tasks, perform well and better than the already used statistical approach. We also observed that MLP underperforms in comparison to the other two techniques. While MLP is capable of learning any non-linear relationship among data features, its performance is greatly depended on the number of epochs and iterations used. Due to the need for the model to be retrained periodically and the time taken by MLP to learn new data; it is not the most viable model for confirmation time prediction. On the other hand, random forest needs to be trained from scratch each time we want the model to learn new data. However, it is observed that it learns new data much faster with lesser complexity as compared to MLP. Random forest, as observed, outperforms all the other models. This is because of random forest model's inherent capability of handling variations and noises in the dataset all the while keeping the model unbiased and stable.

Despite the limitations of MLP and random forest in terms of complexity and training time, machine learning does provide a novel approach when it comes to these confirmation time predictions and can, as observed, outperform statistical methods. One of the future work might be to explore other models that can easily update their knowledge base with new data as well as exploring MLP or other neural structures, for instance Deep Neural Networks, Convolutional Neural Networks etc., to see if changes made to the hyper-parameters, for example, the learning rate or number of units in the hidden layer(s) etc., can decrease relative error magnitude without affecting model accuracy. These changes to hyper-parameters and neural structure may lead to a stronger regression or prediction model which would be more suited to the ever-changing state Ethereum network. Another question that remains would be to understand and explore how the knowledge of time prediction can help the user manipulate the network into making transactions more economic or faster in the future.

## REFERENCES

[1] Blockchains: The great chain of being sure about things, The Economist, http://www.economist.com/briefing/2015/10/31/the-great-chain-of-being-sure-about-things, last accessed 2019/04/12.

[2] Wood, G., Ethereum: A Secure Decentralised Generalised Transaction Ledger, https://gavwood.com/paper.pdf, last accessed 2019/04/12.

[3] Zcash, How it works, https://z.cash/technology/, last accessed 2019/06/04.

[4] CoinPaprika, https://coinpaprika.com/, last accessed 2019/04/10.

[5] Payette, J., Schwager, S., and Murphy, J., Characterizing the Ethereum Address Space, http://cs229.stanford.edu/proj2017/final-reports/5244232.pdf, last accessed 2019/04/10.

[6] Chen, M., Narwal, N. and Schultz, M., Predicting Price Changes in Ethereum, http://cs229.stanford.edu/proj2017/final-reports/5244039.pdf, last accessed 2019/0/11.

[7] Singh H. J., Hafid A. S., Prediction of Transaction Confirmation Time in Ethereum Blockchain using Machine Learning, International Congress on Blockchain and Applications 2019, Avila (Spain).

[8] Eth Gas Station, https://ethgasstation.info, last accessed 2019/04/11.

[9] Greenberg, A. , Crypto Currency https://www.forbes.com/forbes/2011/0509/technology-psilocybin-bitcoins-gavin-andresen-crypto-currency.html#7c44500c353e, last accessed 2019/04/10.

[10] Smart Contracts, https://en.wikipedia.org/wiki/Smart_contract, last accessed 2019/04/12.

[11] Jagers, C. What is Ethereum, https://www.investopedia.com/articles/investing/022516/what-ethereum.asp, last accessed 2019/04/10.

[12] Bajpai, P., Bitcoin Vs Ethereum: Driven by Different Purposes, https://www.investopedia.com/articles/investing/031416/bitcoin-vs-ethereum-driven-different-purposes.asp, last accessed 2019/04/10.

[13] Genesis Block, https://en.bitcoin.it/wiki/Genesis_block 2019/04/12.

[14] Saraf, C. and Sabadra, S., Blockchain platforms: A compendium, IEEE International Conference on Innovative Research and Development (ICIRD) 2018, Bangkok, pp. 1-6 (2018).

[15] Merkle Tree, https://en.wikipedia.org/wiki/Merkle_tree, last accessed 2019/04/12.

[16] Ethereum StackExchange, What's the difference between geth/mist/parity? https://ethereum.stackexchange.com/questions/62653/can-anyone-explain-whats-the-difference-between-mist-geth-parity-in-simple-term, last accessed 2019/07/24.

[17] Ethereum Transaction Information, https://etherscan.io/, last accessed 2019/04/14.

[18] Ethereum Pending Transactions, https://etherscan.io/txsPending, last accessed 2019/04/14.

[19] Ethereum Pending Transaction Queue, https://etherscan.io/chart/pendingtx, last accessed 2019/04/14.

[20] Ethereum, Block Size History, https://etherscan.io/chart/blocksize, last accessed 2019/04/14.

[21] Bootstrap Aggregating, https://en.wikipedia.org/wiki/Bootstrap_aggregating, last accessed 2019/04/14.

[22] Seif, G. Selecting the best Machine Learning algorithm for your regression problem, https://towardsdatascience.com/selecting-the-best-machine-learning-algorithm-for-your-regression-problem-20c330bad4ef, last accessed 04/27/2019.

[23] Arouri, C., Nguifo, E. M., Aridhi, S., Tsopze, N., Towards a constructive multilayer perceptron for regression task using non-parametric clustering. A case study of Photo-Z redshift reconstruction, https://arxiv.org/abs/1412.5513, last accessed 2019/04/14.

[24] Hall, L., Computational Complexity, John Hopkins University, http://www.esi2.us.es/~mbilbao/complexi.htm, last accessed 2019/04/25.

[25] Computational Complexity of Machine Learning Algorithms, The Kernel Trip, https://www.thekerneltrip.com/machine/learning/computational-complexity-learning-algorithms/, last accessed 2019/04/25.

[26] EthGasStation, https://github.com/ethgasstation, last accessed 2019/04/15.

[27] FAQ, EthGasStation, https://ethgasstation.info/FAQcalc.php, last accessed 2019/09/21.

[28] BitInfoCharts, Ethereum (ETH) price stats and information, https://bitinfocharts.com/ethereum/, last accessed 2019/05/25.